\documentclass[12pt,letterpaper]{article}
\usepackage{setspace}
\usepackage[margin=1in]{geometry}
\usepackage[T1]{fontenc}
\usepackage{lmodern}
\usepackage{amsthm,amsmath,amsfonts,amssymb}
\usepackage[authoryear,round]{natbib}
\usepackage{graphicx}
\usepackage{bm}
\usepackage{algorithm}
\usepackage{algpseudocode}
\usepackage{makecell}
\usepackage{booktabs}
\usepackage{xcolor}
\usepackage{xurl}
\usepackage[hidelinks]{hyperref}

\theoremstyle{plain}
\newtheorem{theorem}{Theorem}[section]
\newtheorem{proposition}[theorem]{Proposition}
\newtheorem{corollary}[theorem]{Corollary}
\newtheorem{lemma}[theorem]{Lemma}
\theoremstyle{definition}
\newtheorem{definition}[theorem]{Definition}
\newtheorem{assumption}{Assumption}[section]

\newcommand{\bfnm}[1]{}

\newcommand{\bdoi}[1]{\url{https://doi.org/#1}}

\hypersetup{
  pdftitle={Nonparametric Hypothesis Testing of High-dimensional Clustering With Application to Single-cell RNA Data},
  pdfauthor={Yifan Dai; Di Wu; Yufeng Liu},
  pdfkeywords={Clustering significance; Dimension reduction; Log-concave distributions; Nonparametric statistics; Score-based generative models; Unimodal distributions}
}
\title{\bfseries Nonparametric Hypothesis Testing of High-dimensional Clustering With Application to Single-cell RNA Data}
\author{%
  Yifan Dai\textsuperscript{A}\quad
  Di Wu\textsuperscript{A}\quad
  Yufeng Liu\textsuperscript{B}\\[0.5em]
  {\small\textsuperscript{A}Department of Biostatistics, University of North Carolina}\\
  {\small\textsuperscript{B}Department of Statistics, University of Michigan}\\
  {\small\href{mailto:yufliu@umich.edu}{\texttt{yufliu@umich.edu}}}
}
\date{}
\begin{document}
\maketitle
\begin{abstract}
{Single-cell RNA sequencing studies routinely use clustering to define putative cell types and cell states, yet the observed separation may arise from sampling variability rather than genuine biological heterogeneity. This paper studies formal significance testing of such clustering structure in high-dimensional data. Existing SigClust methods assess clustering significance through Monte Carlo simulation under a Gaussian single-cluster null, but this assumption can be unreliable for normalized gene expression data and other non-Gaussian settings. We propose SigClust-LCP, a nonparametric extension that models a single cluster by a log-concave distribution. To make this approach computationally feasible in moderate to high dimensions, we develop a score-matching estimator for log-concave projection inspired by recent generative modeling ideas. We establish theoretical guarantees for the estimator and for its use in clustering significance testing. Simulations show that SigClust-LCP controls Type-I error more reliably than existing methods across a range of unimodal and mixture distributions while retaining competitive power. In a single-cell RNA sequencing analysis of Hydra cells, the method avoids spurious subclusters within annotated cell populations and supports biologically meaningful separation across lineages and body-axis regions.}
\end{abstract}
\medskip

\section{Introduction}\label{sec:intro}

{Single-cell RNA sequencing (scRNA-seq) has become a central tool for resolving cellular heterogeneity, where clustering is routinely used to define putative cell types and cell states from high-dimensional gene expression profiles. These cluster labels often serve as the starting point for downstream biological interpretation, including marker-gene discovery, trajectory analysis, and benchmarking of computational pipelines. Recent developments in single-cell methodology further underscore the importance of reliable clustering: ClusterDE \citep{songClusterDE2023} addresses false discoveries in post-clustering differential expression, while scDesign3 \citep{songScDesign32024} provides realistic in silico single-cell and spatial omics data for evaluating analysis workflows. Yet a basic question remains largely unaddressed in these pipelines: does an observed partition reflect genuine biological heterogeneity, or can it be explained by a single population with stochastic variation?}

{
In this paper, we focus on the statistical significance aspect of clustering results, motivated by the common situation in scRNA-seq where a clustering algorithm is first applied and the resulting groups are then interpreted as candidate cell populations. In particular, the implementation of classical clustering methodologies such as $k$-means \citep{macqueenMethodsClassificationAnalysis1967} usually requires researchers to specify the desired number of clusters. In practice, the determination of cluster numbers often relies on subjective criteria, such as information-based criterion \citep{sugar2003finding} or examination of cluster stability \citep{von2010clustering}. These methods, however, do not consider the generation model of clusters and cannot provide formal statistical testing for subgroup existence. Consider the case of $k$-means clustering applied to data generated from a single Gaussian distribution: the algorithm clusters the data into distinct groups exhibiting substantial between-cluster separation. Despite evident subgroups suggested by statistics such as the two-sample $t$-test or a stability criterion, such divisions of unimodal data may lack scientific relevance and may lead to erroneous conclusions regarding subgroup existence.}

{Formal statistical testing of clusters have been proposed to evaluate clustering significance for low-dimensional data \citep{walther2002detecting, li2010testing}. These methods adopt a probability interpretation of clusters, assuming a single cluster to be a Gaussian, uniform, or general log-concave distribution. However, tests of mixture distributions, such as likelihood-ratio test \citep{garel2001likelihood} and EM test \citep{li2010testing}, mainly focus on univariate or low-dimensional distributions. Alternative methods utilize Monte Carlo simulation to evaluate clustering results, including the Gap statistic to determine the number of clusters \citep{tibshiraniEstimatingNumberClusters2001} and SigClust to assess the clustering significance \citep{liuStatisticalSignificanceClustering2008}. These methods have been more broadly applied in high-dimensional datasets. Particularly, SigClust has been widely applied to cancer subtype characterization \citep{cancergenomeatlasresearchnetworkComprehensiveGenomicCharacterization2012} and gene regulatory network elucidation \citep{garcia-recioMultiomicsPrimaryMetastatic2023}. In the single-cell setting, the same inferential need arises because clustering decisions affect both downstream differential expression and the evaluation of synthetic or semisynthetic benchmarks.}

\begin{figure}[ht]
    \centering
    \includegraphics[width=0.99\linewidth]{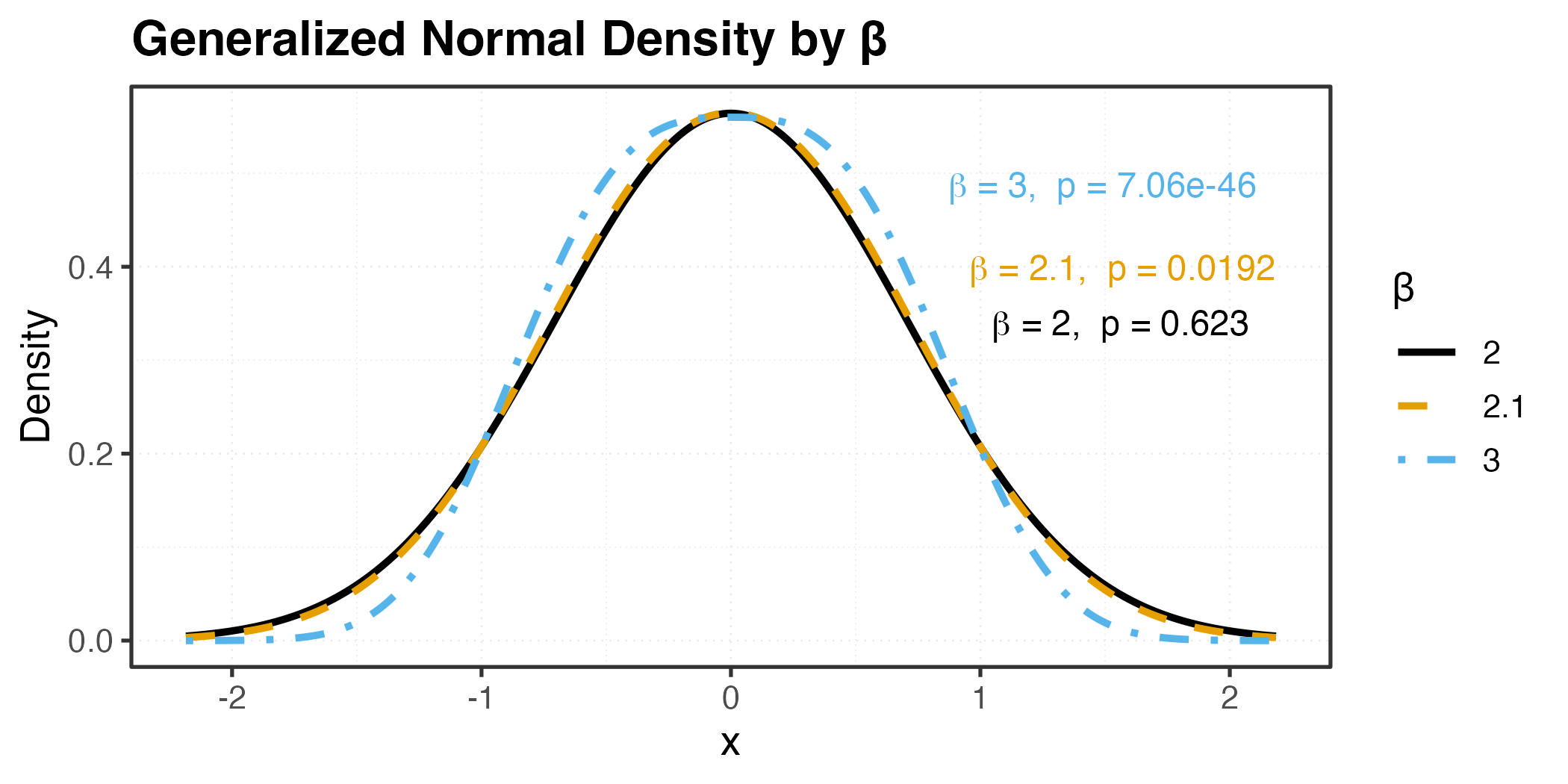}
    \caption{Generalized normal densities and their associated SigClust $p$-values. Here, $\beta$ denotes the shape parameter of generalized normal distributions with $\beta=2$ being standard normal, and the reported $p$-values are median over 10 repetitions using $10,000$ samples.}
    \label{fig:illustration}
\end{figure}

{ Despite the broad applicability of Monte Carlo-based significance of clustering, the methods' fundamental parametric assumption has emerged as a significant limitation in applications such as scRNA-seq. After normalization, transformation, and dimension reduction, gene-expression features may deviate substantially from Gaussianity, even when the underlying population is unimodal. This limitation is evident in Figure \ref{fig:illustration}, which demonstrates elevated Type-I error rates when SigClust is applied to data from a one-dimensional generalized Gaussian distribution with the shape parameter $\beta$. Note that $\beta=2$ corresponds to the standard normal. This family allows for tails that are either heavier than normal (when $\beta<2$) or lighter than normal ($\beta >2$). As shown in Figure \ref{fig:illustration}, despite the clear unimodality of the underlying distribution, SigClust erroneously suggests the presence of multiple clusters. Recent methodological developments have sought to address this limitation through alternative distributional specifications, e.g., \cite{grabskiSignificanceAnalysisClustering2023, DenAdel2025ArtificialVariables}. Other existing methods for evaluating high-dimensional clustering significance, including those developed by \cite{maitraBootstrappingSignificanceCompact2012}, \cite{chakravartiGaussianMixtureClustering2019}, \cite{chenSelectiveInferenceKmeans2023}, \cite{gaoSelectiveInferenceHierarchical2024}, and \cite{dai2025statistical}, remain primarily confined to mixtures of restrictive distributions, leaving the nonparametric test of multi-dimensional clusters an open question.}

{ These considerations motivate a more flexible null model for a single cluster. 
Log-concave distributions consist of a broad class of unimodal distributions with sub-exponential tails, offering more flexibility to model data than Gaussian distributions and have been considered as powerful models for clustering problems, for example, mixtures of log-concave distributions \citep{walther2001multiscale,chang2007clustering, balabdaouiInferenceTwocomponentMixture2018}. We refer the readers to the review of \cite{samworthRecentProgressLogConcave2018} for more details. Log-concave distributions can also be used to test for the presence of mixtures of unimodal distributions. 
In this study, we focus on high-dimensional clustering significance. Analogous to SigClust, we can model a single cluster as data coming from a log-concave distribution and implement a Monte Carlo procedure to obtain the null CI distribution and corresponding $p$-value. Specifically, by estimating the probability distribution function of data via log-concave projection (LCP), the empirical CI distribution under the single cluster assumption can be derived. This enables testing for more general clusters than Gaussian mixtures.}

{A fundamental challenge lies in the estimation of log-concave distributions for high-dimensional data. Although existing literature has established LCP estimators for univariate or low-dimensional settings \citep{dumbgenMaximumLikelihoodEstimation2009, culeLogConcDEADPackageMaximum2009, culeTheoreticalPropertiesLogconcave2010}, its computation can be infeasible for moderate-dimensional data with $d>6$. In particular, the time complexity of local LCP estimators scales polynomially in the dimension $d$, whereas the standard MLE scales exponentially with $d$ \citep{axelrod2019polynomial}. Despite notable empirical speedups for $d\le 6$, the method of \citet{rathke2019fast} still hinges on grid-based numerical integration, so its computational cost can grow sharply with $d$. The lack of computationally efficient estimators can hinder the applicability of log-concave distributions to clustering problems. 
More recently, estimators for a restrictive class within log-concave distributions are available in the literature, as proposed by \cite{samworthIndependentComponentAnalysis2012} with the independent component assumption and \cite{xuHighdimensionalNonparametricDensity2021} with symmetry constraints. However, their assumptions on independency or shape may not hold in practice.}

{ To address this problem, we propose a novel LCP estimator via score matching, a technique widely applied in deep generative models \citep{Hyvarinen2005score, song2021score}, for general high-dimensional data. Unlike MLE-based estimators, the score matching problem seeks minimizing the Fisher divergence between the true distribution and estimated distribution. As a result, the probability score function can be efficiently approximated by a neural network (NN), which provides flexibility for complex density estimation \citep{song2020sliced, song2021score}. This generative-model perspective is also appealing in single-cell analysis, where realistic synthetic data generation has become an important methodological theme \citep{songScDesign32024}. Nevertheless, existing score matching estimators focus on general, non-log-concave distributions and hence cannot remove the clustering structure of the data as LCP does for the null distribution of a single cluster. To fill the gap between general score matching and LCP estimation, we develop a novel constrained score matching method that can be applied to moderate to high dimensional data without shape or independence assumptions. Following the Monte Carlo approach used in SigClust, this estimator further leads to a nonparametric significance test for high-dimensional clusters, namely SigClust-LCP.}

{ The remainder of this paper proceeds as follows: Section \ref{sec:meth} introduces the methodology and computational framework of SigClust-LCP, motivated by the problem of deciding whether putative cellular subpopulations reflect genuine heterogeneity or a single unimodal population. Section \ref{sec:theory} provides analysis of the method's theoretical properties. Section \ref{sec:sim} presents comprehensive comparison of SigClust-LCP with existing methods through simulation studies designed to mimic distributional features encountered in modern biomedical data. Section \ref{sec:real} demonstrates its practical utility through analysis of a high-dimensional single-cell RNA sequencing dataset. We conclude the paper with Section \ref{sec:discussion}, offering discussion of methodological implications and directions for future research.}

\section{Methodology}\label{sec:meth}

{In this section, we develop the methodological ingredients of SigClust-LCP for the applied setting that motivates this paper. SigClust-LCP tests whether a subgroup discovered from high-dimensional data, such as a candidate cell population in single-cell RNA sequencing data, can be explained by a single unimodal distribution. We start with the existing SigClust in Section \ref{sec:meth:sigclust} and some background of LCP in Section \ref{sec:meth:lcp}. Next, we present our proposed nonparametric SigClust-LCP in Section \ref{sec:meth:npsigclust} and describe the score matching LCP estimator used in SigClust-LCP in Section \ref{sec:meth:score}.} 

\subsection{Clustering Significance for Gaussian Mixtures}\label{sec:meth:sigclust}

The original SigClust tests the hypothesis that the data come from a Gaussian distribution against a mixture of Gaussian distributions \citep{liuStatisticalSignificanceClustering2008},  i.e., $H_0: P\sim P_0 \text{ vs.}\ H_1: P\sim \mathcal \alpha P_1+(1-\alpha)P_2$, where $P_0, P_1, P_2$ are Gaussian distributions and $0<\alpha<1$ denotes the Gaussian mixture parameter. To distinguish the alternative hypothesis, SigClust evaluates the clustering separation of the data by 2-means cluster index (CI). The test statistic  CI is defined as the ratio of the within-cluster variation to the overall variation after 2-means clustering,
$$
    CI=\frac{\sum_{a=1}^2\sum_{i\in \mathcal C_a}\lVert \mathbf x_i-{\mathbf {\bar x}^{(a)}}\rVert^2_2}{\sum_{i=1}^{n}\lVert \mathbf x_i-\mathbf{\bar x}\rVert^2_2},
$$ where $\mathbf x_i$ denotes the $i$-th observation, $\mathcal C_a$ is the index set of the $a$-th cluster,  and $\mathbf {\bar x}^{(a)}$ denotes the corresponding within-cluster centroid. Under the null hypothesis, the observations are centered around their mean such that the within-cluster variation is relatively large, resulting in a large CI. Under the alternative hypothesis, the data are well separated, leading to smaller within-cluster variation and consequently smaller CIs.

To formulate formal statistical testing and derive a corresponding $p$-value, SigClust compares the CI for the observed data to the distribution of CIs from the corresponding null distribution, which usually does not have an analytic form. To address this challenge, SigClust adopts a Monte Carlo procedure that iteratively generates $\hat P_0$ as $\mathcal N(\hat \mu, \hat \Sigma)$ and estimates the empirical distribution of CI. Specifically, it first estimates the null data distribution as $\mathcal N(0, {{\hat\Sigma}})$, where the mean component can be set to 0 since CI is location-invariant, and ${{\hat\Sigma}}$ is the estimated covariance matrix of the original data. Then it draws $n$ samples from the null data distribution for $N_{sim}$ times and computes the CI for each batch of generated samples. Finally, the significance of clustering operation is assessed by the empirical $p$-value:
$$
    p=\frac{\#\{CI_m:CI_m\leq CI\}}{N_{sim}},
$$ where $CI_m$ is the cluster index evaluated on the $m$-th batch of generated null samples.

The central assumption of SigClust is that each cluster follows a Gaussian distribution. Although SigClust has been demonstrated its robustness for several non-Gaussian distributions, such as $t$, $\chi^2$, or Poisson distributions \citep{shenStatisticalSignificanceClustering2024}, the Gaussian assumption is still restrictive in many real world applications. As detailed later, we will investigate the conditions when SigClust cannot preserve its Type-I error and becomes anti-conservative. Specifically, if the unimodal data distribution is close to a uniform distribution, the most separated weak-unimodal distribution \citep{tibshiraniEstimatingNumberClusters2001}, the Gaussian reference used in SigClust tends to over-estimate the CI and hence leads to false rejections. This restriction is evident in existing studies investigating the cluster significance for biomedical data, such as single-cell RNA sequencing data \citep{grabskiSignificanceAnalysisClustering2023}. 

\subsection{Log-concave Projection}\label{sec:meth:lcp}

To compare the observed CI with the null CI distribution, SigClust generates the null distribution that (i) preserves the distributional characteristics as much as possible when no clustering structure exists, and (ii) removes the potential clustering structure of observed data.
\cite{maitraBootstrappingSignificanceCompact2012} proposed to randomly rotate the observed data, which preserves the elliptical distributions while breaking non-elliptical distributions, achieving both goals for a wider class of distribution. However, the elliptical assumption can be restrictive and the application primarily focuses on low-dimensional settings. In parallel, the choice of Gaussian null in the original SigClust, despite its parametric restrictions in (i), successfully achieves the goal of breaking clustering structure for high dimensional data. The data generative process of SigClust can be viewed as a \textit{Gaussian Projection} that minimizes the following empirical Kullback–Leibler (KL) divergence
\begin{align*}
    \arg\min_{\mu, \Sigma} \hat{d}_{KL} (f_0, \phi_{\mu, \Sigma}) 
    &= \arg\min_{\mu, \Sigma} \int \log \left( \frac{f_0(\mathbf x)}{\phi_{\mu, \Sigma}(\mathbf x)} \right) \text{d} P_n 
    = \arg\min _{\mu, \Sigma} -\int \log {\phi_{\mu, \Sigma}(\mathbf x)} \text{d} P_n\\
    &=\arg\min_{\mu, \Sigma} \left[ -\log|\Sigma^{-1}| +\text{trace}\left( \Sigma^{-1}(\mathbf X-\mu) (\mathbf X - \mu)^T \right) \right],
\end{align*}
where $\phi_{\mu, \Sigma}$ is the density of Gaussian distribution, $f_0$ is the data density that $\int f_0|\log f_0| < \infty$, $P_n$ is the empirical distribution. Notably, in practice, SigClust penalizes the eigenvalues of $\Sigma$ to ensure the estimability in high-dimensional settings. Upon the estimation of $\hat \mu, \hat\Sigma$, SigClust evaluates the CI on the Monte Carlo samples drawn from the projected Gaussian likelihood $\phi(\hat\mu, \hat\Sigma)$. To achieve nonparametric estimation in (i) for the null distribution, a natural choice is to replace the Gaussian likelihood $\phi_{\mu, \Sigma}$ with a broader class of density functions, such as unimodal distributions. Nevertheless, it is challenging to estimate the general unimodal distributions. Existing nonparametric density estimators mainly focus on general unconstrained distributions instead of unimodal distributions, which fail to achieve the goal of eliminating the potential clustering structure. 

To achieve nonparametric estimation, {the null distribution of a single cluster can be characterized by a log-concave distribution, which is unimodal with exponentially decaying tails by its definition \citep{samworthRecentProgressLogConcave2018}}. A log-concave null distribution thereby breaks any clustering structure in the observed data, naturally satisfying (i) and (ii). The class of log-concave distributions covers a large variety of unimodal distributions, from uniform distribution as the extreme to general or truncated normal distributions, and serves as a flexible choice in modeling real-world data.
More formally, define the class of log-concave densities as $\Phi$, including functions $\phi:\mathbb R^p \to [0, \infty)$ that is log-concave and $\int_{\mathbb R^p} \phi=1.$ The LCP $\hat{\phi}$ is defined as minimizing the empirical KL divergence between $p_0$ and $ \phi \in \Phi$, i.e.,
\begin{align}\label{formula:lcp}
    \hat\phi_{mle}:=\arg\min_{\phi\in \Phi} \hat d_{KL}(p_0, \phi) = \arg\min_{\phi\in \Phi} \int \log\left( \frac{p_0}{\phi}  \right) \text{d} P_n
    = \arg\min_{\phi\in \Phi} -\int \log\phi \text{d} P_n,
\end{align}
which is also an MLE estimator. The computation of LCP has been developed for univariate variables \citep{dumbgenLogcondensComputationsRelated2011} and low-dimensional multivariate variables \citep{culeLogConcDEADPackageMaximum2009}. Notably, the existing algorithms can be computationally feasible in very low dimensions, supporting computation for typically $d \le 6$.

\subsection{Nonparametric Clustering Significance}\label{sec:meth:npsigclust}

To evaluate the clustering significance from non-Gaussian mixtures, we propose SigClust-LCP, a nonparametric clustering significance test. Unlike existing high-dimensional clustering significance test that focuses on multivariate Gaussian clusters, SigClust-LCP generalizes the distribution of a single cluster from Gaussian to the log-concave family $\Phi$. More formally, SigClust-LCP tests the following hypothesis:
\begin{itemize}
    \item $H_0:$ The data follow a log-concave distribution $P\sim \phi$, where $\phi \in \Phi;$
    \item $H_1:$ The data do not follow a log-concave distribution $P\not\sim \phi$ for any $\phi \in \Phi.$
\end{itemize}
Like traditional SigClust methods, SigClust-LCP uses CI as the test statistic to measure the separability of data. To evaluate the distribution of CI under the null $P_{H_0}(CI)$, SigClust-LCP adopts a Monte Carlo approach that first estimates the density of log-concave distribution, and then samples from the estimated density iteratively to compute the empirical distribution of CI. Assuming that the LCP can be efficiently estimated, SigClust-LCP can be summarized in Algorithm \ref{alg:sigclust1}.

\begin{algorithm}[htbp]
\footnotesize
\caption{SigClust-LCP.}\label{alg:sigclust1}
\begin{algorithmic}[1]
\Require{Data matrix $\mathbf{X} \in \mathbb{R}^{n \times d}$, the number of simulations $N_{sim}$}
\Ensure{$p$-value for the two-cluster hypothesis}

\State Apply 2-means clustering on $\mathbf{X}_{n\times d}$ and compute $CI_{\mathbf X}$.

\State Estimate the LCP $\hat\phi_{\mathbf X}$ for $\mathbf X_{n\times d}.$

\For{$i = 1$ to $N_{sim}$}
   \State Generate $\mathbf{Y}^{(i)} \sim \hat\phi_{\mathbf X}$
   \State Apply 2-means clustering on $\mathbf{Y}^{(i)}$ and compute $CI^{(i)}$
\EndFor

\State Compute $p$-value: $p = \frac{1}{N_{sim}}\sum_{i=1}^{N_{sim}} \mathbf{1}(CI^{(i)} \leq CI_{\mathbf X})$

\Return{$p$-value}
\end{algorithmic}
\end{algorithm}

The main challenge for SigClust-LCP is to estimate the null distribution $\hat{\phi}_{\mathbf X}$ via LCP. As mentioned above, the estimation of LCP is challenging for moderate and high dimensional data. In practice, LCP can be computationally costly and inaccurate for $d> 6$, making it intractable to evaluate the clustering significance for high-dimensional data. Therefore, to simplify the problem, instead of assuming that the data come from a $d$-dimensional log-concave distribution, we can assume that the data come from a mixing of $d$ independent log-concave components with unit variance, i.e., 
$
    \mathbf X = \mathbf S A,
$
where $A \in \mathbb R^{d\times d}$, $\mathbf S=[\mathbf s_1,...,\mathbf s_n]^T$, and $\mathbf s_i$ follows an independent $d$-dimensional log-concave distribution with $\text{Cov}(\mathbf s_i)=\mathbf I.$ As a result, the covariance of the observed data is $A^TA.$ To estimate the density of $\mathbf X$, we de-correlate the data by eigen-decomposition of its covariance $\Sigma = A^TA=Q\Lambda Q^T.$ By pre-whitening, we obtain the de-correlated component matrix
$
\mathbf S_o :=  \mathbf X Q\Lambda^{-\frac 12} Q^T=  \mathbf S A Q\Lambda^{-\frac 12} Q^T= \mathbf S O,
$ where $O$ can be shown to be a $d$ by $d$ orthogonal matrix. Therefore, the pre-whitened data $\mathbf S_o$ is an orthogonal projection of the independent components $\mathbf S.$ The estimation of $O$ can be achieved by independent component analysis (ICA) for mixing log-concave distributions \citep{samworthIndependentComponentAnalysis2012}. 
After deriving the ICA representation of the data, we can estimate the original LCP density by estimating the density of $\mathbf S$ dimension-wise. This approach can alleviate the costly computation of MLE-based LCP estimators and hence be applicable to multi-dimensional data.
In practice, we also truncate the eigenvalues $\Lambda$ when we perform pre-whitening, i.e., to obtain $\mathbf S_o$ using top PCs of $\mathbf X$, and perform dimension-wise log-concave density estimation on the reduced PCA space using ICA. The clustering significance is also evaluated on the truncated PCA space. Notably, the PCA space is the same as the classical MDS space so that the space of $\mathbf S_o$ also preserves the pairwise distance and the clustering structure as in SigClust-MDS \citep{shenStatisticalSignificanceClustering2024}.

\subsection{Score Matching Estimation for Log-concave Projection}\label{sec:meth:score}

While the ICA-based LCP alleviates the restriction on dimensionality of the MLE-based LCP, it introduces the independent assumption to the data and ignores the higher-order correlation. To further improve the flexibility of LCP for testing clustering significance, we propose to estimate the gradient of the log-concave log-likelihood $\log \phi$ -- denoted by $\nabla_{\mathbf x} \log \phi$, also known as the score function -- instead of estimating the likelihood directly \citep{Hyvarinen2005score}. Specifically, we define the score matching LCP ${\phi}_{sc}^*$ as following:
\begin{align}
    {\phi}_{sc}^* := \arg \min_{\phi \in \Phi} J(p_0\lVert \phi) := \arg \min_{\phi \in \Phi} \mathbb E\left[
    \left\lVert \nabla_{\mathbf x} \log p_0(\mathbf x) - \nabla_{\mathbf x} \log \phi({\mathbf x}) \right\rVert_2^2
    \right],\label{eq:smlcp}
\end{align}
where $J(p_0\lVert \phi)$ denotes the Fisher divergence between $p_0$, the true data density of the random variable $\mathbf x$, and a log-concave density $\phi$. Under the null hypothesis that $p_0 \in \Phi$, the score matching LCP ${\phi}_{sc}^*$ should be the same as the MLE-based LCP $\phi^*_{mle}$. Under the alternative hypothesis that $f_0 \not \in \Phi$, the score matching LCP $\phi^*_{sc}$ is different from the MLE-based LCP $\phi^*_{mle}$ from MLE, while their connections can be found in de Bruijn's identity and Stein's identity \citep{Hyvarinen2005score}.

Existing literature in score matching provides solutions for Problem (\ref{eq:smlcp}) \citep{vincent2011connection, song2020sliced, song2021score}. Since the data score $\nabla_{\mathbf x} f_0(\mathbf x)$ is not accessible in practice, we consider an alternative target to replace Problem (\ref{eq:smlcp}). Suppose that $\mathbf x$ comes from a latent distribution $\mathbf x'$ such that the conditional density $p(\mathbf x | \mathbf x')$ is known (e.g., a standard Gaussian distribution). To estimate the gradient of log-likelihood of $\nabla_{\mathbf x} \log p_0(\mathbf x)$ at some point $\mathbf x_0$, it suffices to estimate the expectation of the gradient of $p(\mathbf x_0|\mathbf x')$ with respect to $\mathbf x'$. This observation leads to the denoising score matching problem defined as follows.
\begin{definition}
    Define denoising score matching problem as
    \begin{align}
        {\phi}_{noise}^* := \arg\min_{\phi \in \Phi} \mathbb E_{\mathbf x, \mathbf x'} \left[ 
        \left\lVert
        \nabla_{\mathbf x} \log p(\mathbf x | \mathbf x') - \nabla_{\mathbf x} \log \phi(\mathbf x)
        \right\rVert_2^2
        \right],\label{eq:dsm}
    \end{align}
    where $\mathbf x \sim p_0$ as in (\ref{eq:smlcp}), and $p(\mathbf x|\mathbf x')$ is the conditional density of $\mathbf x$ given a random $\mathbf x'$.
\end{definition}
Compared to Problem (\ref{eq:smlcp}), the denoising score matching LCP estimates the gradient of a conditional probability $p(\mathbf x|\mathbf x'),$ which can be set to a Gaussian distribution with known covariance by setting $\mathbf x = \mathbf x' + \sigma \mathbf z$, where $\mathbf z \sim \mathcal N(0, \mathbf I)$. In this case, Problem (\ref{eq:dsm}) reduces to 
\begin{equation}\label{eq:noise}
\begin{aligned}
    {\phi}_{noise}^*
    &= \arg \min_{\phi \in \Phi}\mathbb E_{\mathbf x, \mathbf z}\left[\left \lVert \nabla_{\mathbf x} \log \phi(\mathbf x)+\frac{\mathbf z}{\sigma} \right\rVert^2_2\right] \\
    &= \arg \min_{\phi \in \Phi}\mathbb E_{\mathbf x', \mathbf z}\left[\left \lVert \nabla_{\mathbf x} \log \phi(\mathbf x' + \sigma \mathbf z)+\frac{\mathbf z}{\sigma} \right\rVert^2_2\right].
\end{aligned}
\end{equation}
The denoising score matching provides a convenient approach to estimate the gradient of likelihood by adding random noise to the observed sample $\mathbf x'$ and then approximating the added noise using the gradient $\nabla_{\mathbf x} \phi(\mathbf x)$, which can be intuitively understood as a denoising process.
\begin{proposition}\label{prop:dsm} \citep{vincent2011connection}
    For up to a constant C independent of $\phi$, it holds that
    \begin{align*}
    \mathbb E_{\mathbf x, \mathbf x'} \left[ 
        \left\lVert
        \nabla_{\mathbf x} \log p(\mathbf x | \mathbf x') - \nabla_{\mathbf x} \log \phi(\mathbf x)
        \right\rVert_2^2
        \right] =
        \mathbb E\left[
    \left\lVert \nabla_{\mathbf x} \log p_0(\mathbf x) - \nabla_{\mathbf x} \log \phi({\mathbf x}) \right\rVert_2^2
    \right] + C
    \end{align*}
    for any $\phi \in \Phi$.
\end{proposition}
Proposition \ref{prop:dsm} shows that the denoising score matching LCP is the same as the standard score matching LCP, i.e., $\phi^*_{noise} = \phi^*_{sc}$. Due to the equivalence, we do not distinguish the standard score matching with the denoising score matching for the rest of this paper. Since Problem (\ref{eq:noise}) no longer involves non-accessible terms, we can derive a sample-based score matching LCP as following:
\begin{equation}
    \hat{\phi}_{sc}^* := \arg \min_{\phi \in \Phi} \sum_{i=1}^n  \sum_{j=1}^m
    \left\lVert
    \nabla_{\mathbf x} \log \phi(\mathbf x_i+\sigma \mathbf z_{ij}) + \frac{\mathbf z_{ij}}{\sigma}
    \right\rVert^2_2,
\end{equation}
where $m$ is the number of different normal noise $\mathbf z_{ij} \sim\mathcal N(0, \mathbf I)$ added to $\mathbf x_i$'s, typically large.

To solve the sample score matching problem, we can use a neural network (NN) to approximate the log-likelihood function $\log \phi(\mathbf x)$ -- denoted as $\psi(\mathbf x; \bm\theta)$, where $\bm\theta \in \Theta$ denotes the parameters associated with the NN. Specifically, the family of NNs of $L$ layers $\mathcal F_L$ can be represented as the following:
$
\psi(\mathbf x; \Theta)=\mathcal L_1\circ o_1 \circ \cdots \circ \mathcal L_{L-1}\circ o_{L-1} \circ \mathcal L_L, 
$
where $\mathcal L_i(\mathbf x) = W_i \mathbf x +b_i$ is a multi-variate linear function for $i=1, \dots, L$, and $o_i(\mathbf x)$ is an element-wise nonlinear activation function \citep{lecun2015deep}. While NN provides flexibility to estimate the usually nonlinear and complex density function, it cannot guarantee that the estimated score function comes from a log-concave distribution. Hence, a critical problem is to ensure that $\psi$ falls in the class of concave functions, denoted as $\Psi:=\left\{f:\mathbb R^d \to \mathbb R \text{ such that } \int_{\mathbb R^d} e^{f}=1 \text{ and concave}\right\}$. A sufficient condition for $\psi$ to be concave is to have (i) the elements of the weight matrices (i.e., $W_i$'s) from linear layers $\mathcal L_2, \dots, \mathcal L_{L-1}$ to be non-negative, (ii) the elements of the weight matrix $W_L$ from the last linear layer $\mathcal L_L$ to be non-positive, and (iii) all activation functions $o_1, \dots, o_L$ are convex. Once the three conditions are met, the resulting neural network can be shown to be concave. This follows from the fact that the composition of a convex function and a convex non-decreasing function is also convex \citep{amos2017input}. Consequently, taking the negative of a convex function yields a concave function. Denote the family of input concave networks described above as $\Psi_{ICN}$, and the final estimator of the log-concave gradient is given by:
\begin{equation}\label{eq:edsm}
    \hat\psi_{\Theta} := \arg \min_{\psi \in \Psi_{ICN}} \sum_{i=1}^n  \sum_{j=1}^m
    \left\lVert
    \nabla_{\mathbf x} \psi(\mathbf x_i + \sigma \mathbf z_{ij};\bm \theta) + \frac{\mathbf z_{ij}}{\sigma}
    \right\rVert^2_2.
\end{equation}
Note that the NN-fitted log-likelihood $\hat\psi_{\Theta}$ can be unrestricted and hence may not be unique, but its gradient and corresponding probability density $\hat{\phi}_{\Theta}$ are uniquely determined. In practice, $\sigma$ can be empirically set to 0.1 (preferably a small value), and the specification of NN hyper-parameters and optimizers can be found in the Supplementary Material and is investigated in the simulation studies.

Finally, if the score matching LCP from (\ref{eq:smlcp}) is accurately estimated, the samples can be iteratively generated from the unadjusted Langevin algorithm (ULA) \citep{dalalyan2017theoretical, song2021score}:
$
\mathbf x_{t} = \mathbf x_{t-1} - \tau \nabla_{\mathbf x} \log\phi(\mathbf x) +\sqrt {2\tau} \mathbf z, \ \mathbf z\sim \mathcal N(0, \mathbf I),\ t=1,..,K,
$ where $\tau$ is a step parameter, and $T$ is the number of iterations. As described above, we can approximate the null distribution of CI by generating $N_{sim}$ batches of Monte Carlo samples. The corresponding $p$-value can be derived from the quantile of the empirical CI distribution determined by the observed CI. SigClust-LCP using score matching can be summarized in Algorithm \ref{alg:sigclust2}.

\begin{algorithm}[t]
\footnotesize
\caption{SigClust-LCP(Score): clustering significance via score matching.}\label{alg:sigclust2}
\begin{algorithmic}[1]
\Require{Data matrix $\mathbf{X} \in \mathbb{R}^{n \times d}$, the number of simulations $N_{sim}$, ULA step size $\tau$, ULA iterations $K$}
\Ensure{$p$-value for the two-cluster hypothesis}

\State Apply 2-means clustering on $\mathbf{X}_{n\times d}$ and compute $CI_{\mathbf X}$

\State Estimate the log-concave gradient of $\hat{\psi}_{\Theta}$ using $\mathbf X$

\For{$j = 1$ to $N_{sim}$}
    \State Sample $\{\mathbf x_{0, i}^{(j)}\}_{i=1:n} \sim \mathcal U(-2, 2)^d$
    \For{$t=1$ to $K$}
    \State Update 
    $\mathbf x^{(j)}_{t,i} = \mathbf x^{(j)}_{t-1,i} 
    - \tau \nabla_{\mathbf x} \hat{\psi}_{\Theta}\!\left(\mathbf x^{(j)}_{t-1,i}\right) 
    + \sqrt{2\tau}\,\mathbf z_{t,i}, \quad i = 1,\dots,n,$ 
    where $\mathbf z_{t,i} \sim \mathcal{N}(0,1)$ are independent Gaussian random variables, independent of all other sources of randomness.
    \EndFor
    \State Update $\mathbf X^{(j)} = [\mathbf x^{(j)T}_{K,1}, \cdots, \mathbf x^{(j)T}_{K,n}]^T$
   \State Apply 2-means clustering on $\mathbf{X}^{(j)}$ and compute $CI^{(j)}$
\EndFor

\State Compute $p$-value: $p = \frac{1}{N_{sim}}\sum_{j=1}^{N_{sim}} \mathbf{1}(CI^{(j)} \leq CI_{\mathbf X})$

\Return{$p$-value}
\end{algorithmic}
\end{algorithm}

\section{Theoretical Results}\label{sec:theory}

In this section, we first establish the asymptotic properties for proposed score matching LCP in Section \ref{sec:theo:lcp}. We then extend the oracle properties of SigClust from the Gaussian setting to the broader log-concave family in Section \ref{sec:theo:sigclust}, showing that SigClust preserves Type I error control and remains powerful when the reference distribution for Monte Carlo simulation is known. We further provide theoretical guarantees when the reference distribution is estimated from the observed data. Finally, we investigate the empirical high-dimensional scenario with the dimension reduction in Section \ref{sec:theo:pca}. Specifically, we analyze the asymptotic behavior of $p$-values for SigClust-LCP when PCA is applied to reduce the data dimension.

\subsection{Asymptotic Properties for Score Matching LCP}\label{sec:theo:lcp}

One important question of score matching LCP is whether the estimated LCP is consistent under reasonable assumptions. We consider the scenario when the log-likelihood of target LCP $\psi$ belongs to the sub-family that can be fitted by the concave NN indexed by $\bm\theta$, denoted as $\Psi_{\Theta}:=\{\psi(\cdot, \theta):\mathbb R^d \to \mathbb R, \bm\theta \in \Theta\}$. We assume that the space $\Theta$ is compact in $\mathbb R^\tau$. Moreover, we consider the Lipschitz continuous score functions with respect to $\theta\in \Theta$. With some regularity conditions on the log-concave density $\phi$, we can show the asymptotic consistency between $\hat{\phi}_{sc}$ (the corresponding density of the unconstrained log-likelihood $\hat{\psi}_{sc}$) by solving Problem (\ref{eq:edsm}) and $\phi^*_{sc}$ in terms of the Fisher divergence $J(p_0\lVert \cdot).$

\begin{assumption}\label{a:compact}
    (Compactness) The parameter space $\Theta$ is compact in $\mathbb R^\tau$.
\end{assumption}

\begin{assumption}\label{a:continuity}
    (Lipschitz continuity) The function $\nabla_{\mathbf x} \log \phi(\cdot; \theta)$ is Lipschitz continuous with respect to $\theta \in \Theta$ in terms of Frobenious norm. Specifically, for any $\theta_1, \theta_2 \in \Theta$, we define $\phi_1 := \phi(\cdot; \theta_1) \in \Phi_{\Theta}$ and $\phi_2:= \phi(\cdot; \theta_2)\in \Phi_{\Theta}$, and they satisfy $\lVert \nabla_{\mathbf x}\log \phi_1(\mathbf x) - \nabla_{\mathbf x} \log \phi_2(\mathbf x)\rVert_F \le L_{\phi}(\mathbf x)\lVert \theta_1-\theta_2\rVert_2,$ with $\mathbb E[L_{\phi}^{2(1+\delta)}(\mathbf x)] <\infty$ for some $\delta >0.$
\end{assumption}

\begin{assumption}\label{a:polynomial}
    (Uniformly polynomial growth) There exists $r\ge 0$ such that
    $ \lVert \nabla_{\mathbf x} \log \phi(\mathbf x)\rVert_2 \lesssim 1 + \lVert \mathbf x\rVert_2^r$ uniformly over $\Phi_{\Theta}$ for any $\mathbf x \in \mathbb R^d.$
\end{assumption}

\begin{theorem}
    Assume that the conditions discussed above are satisfied. Let $\hat{\phi}_{sc}$ be the density of the unconstrained log-likelihood $\hat{\psi}_{sc}$ by solving Problem (\ref{eq:edsm}) and $\phi^*_{\Theta}$ be the restricted score matching LCP with $\log \phi^*_{\Theta} \in \Psi_{\Theta}.$ Then for every $n\in \mathbb N^+$, with probability at least $1-\delta$, we have $J(p_0\lVert\hat{\phi}_{\Theta}) - J(p_0\lVert{\phi}_{\Theta}^*)  \lesssim  \frac{\text{diam}(\Theta)}{\delta} \sqrt{\frac \tau n}.$
\end{theorem}

Assumptions \ref{a:compact}, \ref{a:continuity}, and \ref{a:polynomial} ensure that the family of probability densities should be smooth enough and satisfy a weaker condition of strongly log-concavity. 
Typically, the number of parameters \(\tau\) of an NN class increases in linear or polynomial order of the data dimension \(d.\) Consequently, the excess risk bound shows that the score matching LCP can accurately estimate the log-concave density for a polynomial sample complexity to the data dimension, i.e., \(n=O(d^\beta), \beta >1\).

\subsection{Properties of SigClust-LCP}\label{sec:theo:sigclust}

Theorem 3.1 of \cite{shenStatisticalSignificanceClustering2024} proves the convergence of CI in one-dimensional symmetric, continuous settings. In such cases, the solution to the population $k$-means problem is $\mathbb E|X|.$ Therefore, the observed CI converges to
\begin{equation*}
    CI_{X} \xrightarrow{a.s.} \frac{\mathbb E[X^2] -(\mathbb E|X|)^2}{\mathbb E[X^2]}=1-\frac{(\mathbb E|X|)^2}{\mathbb E[X^2]}.
\end{equation*}
However, in practice, SigClust specifies $X\sim \mathcal N(0, \hat\sigma^2),$ where $\hat\sigma^2 = \frac 1n \sum_{i=1}^n X_i^2 \xrightarrow{a.s.} E[X^2].$ Therefore, the Monte-Carlo null CI of SigClust in one run is given by
\begin{align*}
    CI_{null} = 1 - \frac{\sum_{i=1}^n |Y_i\hat\sigma|^2}{\sum_{i=1}^n Y_i^2 \hat\sigma^2} =1-\frac{\sum_{i=1}^n |Y_i|^2}{\sum_{i=1}^n Y_i^2} \xrightarrow{a.s.}  1-\frac{(\mathbb E|Y|)^2}{\mathbb E[Y^2]}=1 - \frac{2}{\pi},
\end{align*}
where $Y_i$ and $Y$ are drawn from a standard normal distribution $\mathcal N(0, 1)$ \citep{huangStatisticalSignificanceClustering2015}. Notably, in one-dimensional cases, SigClust does not need to estimate the data variance $\hat{\sigma}^2$ since CI is invariant to scaling. Therefore, under the null hypothesis of unimodality instead of Gaussianity, the null CI of the original SigClust is biased with the empirical $p$-value converging to $0$ in probability when the observed data satisfy
$R_{F_X}:=\frac{(\mathbb E|X|)^2}{\mathbb E[X^2]} > \frac{2}{\pi},$ failing to preserve the Type-I error in such cases. Conversely, the original SigClust can be conservative when $R_{F_X}< 2/\pi,$ i.e., the $p$-value converges to 1 in probability. 

The discussion under the one-dimensional symmetric, continuous settings - despite its highly restrictive assumptions - provides guidance of applying SigClust to practical problems. For example, the generalized normal distributions and beta distributions usually have $R_{F_X}> \frac{2}{\pi}$, with anti-conservative SigClust $p$-values. Meanwhile, the original SigClust is still valid while conservative for a large family of distributions, including student's $t$ distribution with $R_{F_X}<\frac{2}{\pi}$ for finite degrees of freedom. The following theorem and corollary extend the one-dimensional scenario to multivariate cases.

\begin{theorem}\label{theorem1}
Assume that the random variable $\mathbf x$ is independently generated from some distribution $F(\cdot)$ with continuous density $f(\cdot).$ Suppose that the density function is symmetric and each variable is independent to each other. Assume that $\mathbb E[|\mathbf x_{.1}|] \geq \cdots \geq \mathbb E[|\mathbf x_{.d}|].$ Then for $\mathbf x\sim F$, we have
$
CI \xrightarrow{a.s.} 1 - \frac {\left( \mathbb E[|\mathbf x_{.1}|] \right)^2} {\mathbb E[\lVert \mathbf x\rVert_2^2]}.
$
\end{theorem}

\begin{corollary}
    Suppose that the assumptions of Theorem \ref{theorem1} holds, and 
    $
    \frac {\left( \mathbb E[|\mathbf x_{.1}|] \right)^2} {\mathbb E[\lVert \mathbf x\rVert_2^2]} > \frac{2}{\pi}\frac{\sigma_1^2}{\sum_{j=1}^d \sigma_j^2},
    $
    where $\sigma_1^2,...,\sigma_d^2$ denote the marginal variances of different dimensions, then empirical $p$-value from the original SigClust converges to $0$ in probability as $n\to \infty.$
\end{corollary}

Next, we move forward to prove the preservation of Type-I error under the settings of log-concave distributions. The result shows that the statistical size can be controlled if the log-concave distribution $\phi_{\mathbf x}$ is known.

\begin{theorem}\label{theorem2}
    Suppose that the data $\mathbf x_1,...,\mathbf x_n$ follow a log-concave distribution $\phi_{\mathbf x}$, then the empirical $p$-values of SigClust-LCP follow a $U(0, 1)$ distribution.
\end{theorem}

In practice, the reference distribution $ \phi_{\mathbf x}$ is estimated from data, and the impact of its convergence rate on $p$-value consistency is important. However, to the best of our knowledge, existing literature has not investigated this case but generally assume that the null distribution is known \citep{tibshiraniEstimatingNumberClusters2001, liuStatisticalSignificanceClustering2008, chakravartiGaussianMixtureClustering2019, shenStatisticalSignificanceClustering2024}. Therefore, we provide a general theorem to investigate this problem.

\begin{theorem}[$m$ out of $n$ significance test]\label{theorem:consistency}
    Under regularity assumptions described in Supplementary Materials, let $\mathbf x_{1:n}\sim P^n$ and $Q_n$ be any reference law on $\mathbb R^d$. For $m\ge 4$,
    \begin{align*}
    \sup_{\alpha \in [0, 1]} \left| \mathbb P_P\{p_m(Q_n)\le \alpha\} - \alpha\right|
    &\lesssim \sqrt{m^{\frac 12}\left( W_2(P,Q_n)^2+W_2(P,Q_n)\right)} \\
    &\quad + W_2(P,Q_n)^2+W_2(P,Q_n),
    \end{align*}
    where $p_m(Q_n)$ denotes the $p$-value using $Q_n$ with $m$ observations, and $W_2(P, Q)$ is the Wasserstein-2 distance between $P$ and $Q$.
\end{theorem}

To achieve uniformly consistent testing, we consider three scenarios: (i) if $m<\infty$, then it suffices to have $W_2(P, Q_n) \overset{p}{\to} 0,$ which is satisfied by most empirical estimators; (ii) if $m\to \infty$, $n \to \infty$, with $n/m \to \infty$, then a sufficient condition for uniform convergence would be that $m^{1/2}W_2(P, Q_n) \overset{p}{\to} 0;$ (iii) otherwise, if $n/m=O(1)$, then we require $W_2(P, Q_n) = o_P(n^{-1/2}),$ which can be hardly achieved in practice. Specifically, most parametric estimators (e.g., sample mean and variance for $1d$-normal distributions) lead to $W_2(P, Q_n)=O_P(n^{-1/2})$, a rate failing to achieve the consistency of $p$-values.

Theorem \ref{theorem:consistency} suggests that testing using the complete data with very large $n$ can lead to inconsistent $p$-values even under the Gaussian assumption. This result aligns with numerical experiments for large sample sizes, where the empirical distribution of original SigClust $p$-values converges to a S-shaped curve instead of the expected diagonal line in $[0, 1]$ (see Supplementary Materials). As a remedy, we propose to test subsets of the data for large datasets -- with each $p$-value consistent -- and then aggregates these $p$-values to boost the power. Details for the improved algorithm can be found in Supplementary Materials.

We proceed to present the asymptotic power of SigClust-LCP for finite sample size $n$ and infinite data dimension $d$. The result is similar to that in \cite{liuStatisticalSignificanceClustering2008}, which characterizes the $p$-values for independent Gaussian mixtures. However, our result waives the independent Gaussian mixture assumption made by \cite{liuStatisticalSignificanceClustering2008} and shows that all SigClust methods -- including the original Gaussian SigClust -- are powerful for mixtures of any distributions satisfying mild conditions in moment and covariance described in the  theorem below. This includes non-Gaussian distributions with higher order correlations.

\begin{theorem}\label{theorem3}
    Suppose that the data $\mathbf x_1, ..., \mathbf x_n$ are independent samples from two clusters $C_1$ and $C_2$ with
    $
    \mathbf x_i = \delta I(i \in C_1) \mathbf 1_d + \bm \epsilon_i,
    $
    where $\bm \epsilon_i$ follows a log-concave distribution $\phi_{\bm \epsilon}$ with $E[\bm \epsilon_i] = 0$ and $E[\bm \epsilon_i^T \bm \epsilon_i] = \Sigma_{\bm \epsilon}.$ 
    Assume that 
    (i) 
    $\frac{
        \sum_{j=1}^d \lambda_{j, \bm \epsilon}^2
        }{
        \text{Tr}(\Sigma_{\bm \epsilon})^2
        } \to 0$ as $d\to \infty$, where $\lambda_{j, \bm \epsilon}$'s are the eigenvalues, 
    and (ii) 
    $ \text{Tr}(\Sigma_{\bm \epsilon}) = O(d^\gamma)$ with $\gamma \in [0, 1).$
        If $\phi_{\epsilon}$ is known, then for a fixed $n$, 
    the empirical $p$-value from Algorithm \ref{alg:sigclust1} converges to 0 in probability as $d \to\infty$.
\end{theorem}

\subsection{Property for PCA-based Dimension Reduction}\label{sec:theo:pca}

In practice, to achieve good performance for very high-dimensional data, SigClust is recommended to be used along with PCA or MDS \citep{shenStatisticalSignificanceClustering2024}, reducing the dimension from $d$ to $q$, where $q \ll d$. In this section, we justify the use of PCA with SigClust-LCP by its property of low rank recovery. Specifically, consider a factor model with
$\mathbf X = \mathbf c\bm \delta^T + \Xi = \mathbf M + \Xi,$
where $\mathbf c = (\mathbf 1_{n_1}^T, -\mathbf 1_{n_2}^T)^T$ is the vector of cluster assignment, $\bm \delta \in \mathbb R^d$ is the vector of mean shifting, and $\Xi$ is an $n\times d$ mean-zero perturbation matrix. It is known that the leading PC of $\mathbf X$ can recover that of $\mathbf M$ when rows of $\Xi$ are independent sub-Gaussian vectors and the signal-to-noise ratio (SNR), defined as $\frac{\lVert \delta \rVert^2}{\max_{i, j} \rVert \Phi_{ij}\lVert_{\psi_2}^2}$, is large enough \citep{littleAnalysisClassicalMultidimensional2023, shenStatisticalSignificanceClustering2024}. Notably, the leading PC of the mean matrix $\mathbf M$ is proportional to the vector of cluster assignment $\mathbf c$, which has a CI of 0. However, the sub-Gaussian assumption may not hold if entries of $\Xi$ follow log-concave distributions, which is sub-exponential \citep{samworthRecentProgressLogConcave2018}. 
Nonetheless, the next lemma states that, for a mixing of independent sub-exponential distributions, given the SNR of clustering effects similar to that for sub-Gaussian scenarios, the leading PC of the observed data can still recover the clustering structure.

\begin{lemma}\label{lemma}
    Suppose that the data $\mathbf x_1, ..., \mathbf x_n$ are independent samples from two clusters $C_1$ and $C_2$ with
    $
    \mathbf x_i = \bm\delta [2I(i \in C_1)-1]  + A^T\bm \epsilon_i,
    $
    where $\bm \epsilon_i$'s follows independent log-concave distributions $\phi_{\bm \epsilon_i}=\prod_{k=1}^d\phi_{\epsilon_{ik}}$ with $\mathbb E[\bm \epsilon_i] = 0$ and $\text{Var}[A^T\bm \epsilon_i] = A^TA =\Sigma$ for all $i=1, \cdots , n$. Assume that $\frac{\lVert \bm \delta\rVert^2}{\sigma_{\max}^2 \lambda_1 } = \omega \left(\sqrt d \log n +(\log n)^2 + d/n\right),$ where $\sigma_{\max}:= \max_{i, k} \lVert \epsilon_{ik}\rVert_{\psi_1},$ and $\lambda_1$ is the largest eigenvalue of $\Sigma.$ Then we have $\lVert \tilde{\mathbf{pc}}_1- \tilde{\mathbf c}\rVert_{\infty} =o_P(\frac{1}{\sqrt n}),$ where $ \tilde{\mathbf{pc}}_1:=\frac{\hat{\mathbf{pc}}_1}{\lambda_1}$, $\hat{\mathbf {pc}}_1 $ is the estimated leading PC, and $\tilde{\mathbf c}=\frac {1} {\sqrt n}(\mathbf 1_{n_1}^T, -\mathbf 1_{n_2}^T)^T$.
\end{lemma}

{The definition of SNR in Lemma \ref{lemma} is largely comparable and bounded by an absolute constant to that of existing studies, although we slightly adjust it to accomodate the independent component assumption. In comparison, \cite{littleAnalysisClassicalMultidimensional2023} presented a rate of $\omega( \sqrt d \log n + d/n)$ for the SNR of sub-Gaussian variables. In the low-dimensional regime, the convergence of the leading PC requires $\text{SNR}=\omega(\log n)$ for sub-Gaussian variables and $\text{SNR}=\omega(\log^2 n)$ for mixing sub-exponential distributions. In the high-dimensional regime with $d \ge O(\log^2 n)$, the rates are similar for sub-Gaussian and sub-exponential distributions.}

After obtaining the leading PC that converges to two points representing cluster assignment with a CI of 0, the power of SigClust can be analyzed by lower bounding the null CI. Specifically, we consider least favorable one-dimensional unimodal distribution that minimizes CI. Theorem 1 in \cite{tibshiraniEstimatingNumberClusters2001} shows that CI of all unimodal distributions cannot be larger than that of a uniform distribution with a CI of $0.75$. Since the observed CI converges to 0 while the null CI is bounded away from 0, the next theorem demonstrates the power of SigClust-LCP.

\begin{theorem}
    Suppose that the settings and assumptions in Lemma \ref{lemma} hold.
    The empirical $p$-value with $q=1$ converges to 0 in probability.
\end{theorem}

\section{Simulation Studies}\label{sec:sim}

{In this section, we evaluate the performance of SigClust-LCP across a range of simulated distributions and compare its behavior with other SigClust methods, $k$-means inference \citep{chenSelectiveInferenceKmeans2023}, RIFT, and MRIFT \citep{chakravartiGaussianMixtureClustering2019}. We also compare with baselines specifically designed for scRNA-seq data, including scSHC \citep{grabskiSignificanceAnalysisClustering2023}, CHOIR \citep{Sant2025CHOIR}, and Recall \citep{DenAdel2025ArtificialVariables}. The simulation settings are chosen to mimic the distributional irregularities that can arise in transformed and dimension-reduced single-cell RNA sequencing data, where a single cell population may still be markedly non-Gaussian. Specifically, we consider two classes of distributions. The first class includes unimodal distributions, which serve as the null hypothesis for SigClust-LCP, and the second class consists of mixtures of two distinct unimodal distributions, representing the alternative hypothesis that SigClust-LCP should reject the null. We assess performance by examining the empirical cumulative distribution function (CDF) of $p$-values over 100 repetitions. Under the null hypothesis, a valid test should yield an empirical CDF that closely approximates a uniform distribution over the interval \([0,1]\) (i.e., \(\mathcal{U}(0,1)\)), whereas under the alternative hypothesis the $p$-values should be concentrated near 0.}

The unimodal reference distributions consist of (i) continuous distributions, including the generalized normal distribution, the beta distribution, and the standard normal distribution; and (ii) count distributions: negative binomial distributions and multinomial distributions. 
Accordingly, we perform centering for (i) and log-normalization for (ii), project the data onto the first 50 principal components for SigClust-LCP and the comparison methods. For scSHC, SigClust-MDS, and ICA-based SigClust-LCP, we use their respective proposed normalization and dimension-reduction procedures.


\subsection{Generalized Normal Distributions}

We first evaluate the performance using generalized normal distributions, whose density is given by
\(
f(x)=\frac{\beta}{2\alpha\,\Gamma(1/\beta)} \exp\left\{-\left(\frac{|x-\mu|}{\alpha}\right)^\beta\right\},
\)
where \(\alpha\) is the scale parameter, \(\mu\) is the location (mean) parameter, and \(\beta\) controls the tail behavior. This distribution has a unique mode at \(\mu\). Note that in the univariate case, if \(\beta > 2\), the original SigClust method is anti-conservative and fails to control its Type-I error rate. In our experiments, we generate \(d\)-dimensional independent, unit-variance generalized normal variables with \(\beta=4\). The mean vector \(\boldsymbol{\mu}=(\mu_1, \mu_2, \dots, \mu_d)^T\) is initially drawn from a multivariate normal distribution \(\mathcal{N}(0, \mathbf{I}_d)\) and then fixed for half of the samples, while for the remaining samples it is further perturbed by adding noise from \(\mathcal{N}(0,a)\) in one variable. Setting \(a=0\) corresponds to the null hypothesis, in which no cluster structure is present in the data. 

\begin{figure}[ht]
    \centering
    \includegraphics[width=0.99\linewidth]{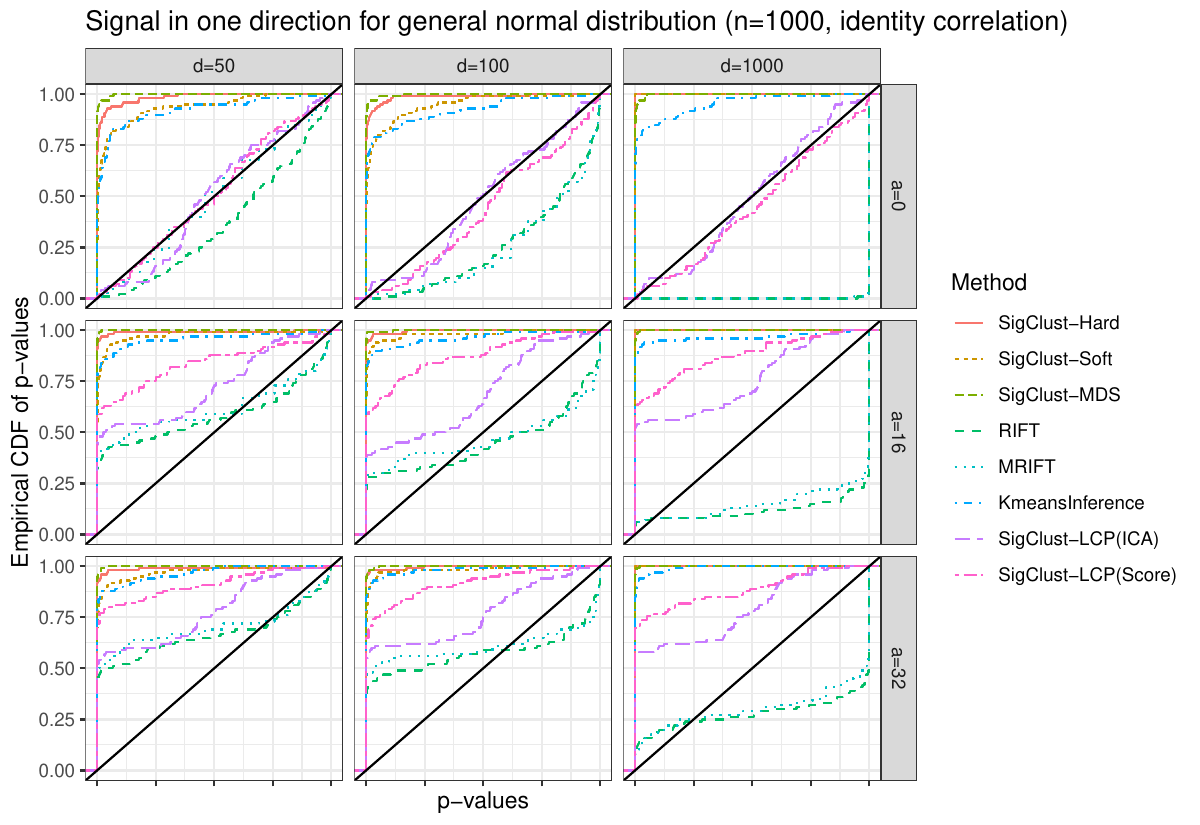}
    \caption{Simulation results for generalized normal distributions. Here, $a$ denotes the distance between two clusters and $a=0$ denotes the null hypothesis of a single cluster.}
    \label{fig:sim_gnorm}
\end{figure}

Figure~\ref{fig:sim_gnorm} presents a comparison of the performance of SigClust-LCP with other methods under the settings of \(n=1000\), a unidirectional signal, and independent variables. As expected, SigClust-LCP, RIFT, and MRIFT are the only methods that consistently maintain their Type-I error at the nominal 0.05 level across all simulation settings. In contrast, all other methods fail to control their sizes for all three dimensional settings. Regarding statistical power, we observe that SigClust-LCP is comparable to RIFT and M-RIFT when $d=50$; they become more likely to reject the null hypothesis as the cluster difference increases. As the dimension increases, SigClust-LCP based on the score matching estimator, denoted as SigClust-LCP(Score), shows advantages in power over ICA-based SigClust-LCP (denoted as SigClust-LCP(ICA) in Figure~\ref{fig:sim_gnorm}), which further outperforms RIFT and MRIFT. Other methods are not directly comparable due to their loss of size control. Additional results for other scenarios, including different sample sizes, varying signal strengths, and alternative correlation structures, are provided in the Supplementary Materials. Notably, SigClust-LCP is robust to non-independent correlations, preserving its Type-I error under the null while remaining powerful under the alternative, while RIFT and M-RIFT are sensitive to dependent data.

\subsection{Beta Distributions}
The Beta distribution is a class of bounded continuous distributions with the density function
\(
f(x)=\frac{x^{\alpha-1}(1-x)^{\beta-1}}{\mathrm{B}(\alpha, \beta)},\quad x\in[0, 1],
\)
where \(\alpha\) and \(\beta\) are shape parameters and \(\mathrm{B}(\alpha, \beta)\) denotes the Beta function. Note that the Beta distribution is not always unimodal (for example, when \(\alpha=\beta=0.5\)); in this study, we constrain it to be strictly unimodal by setting \(\alpha=\beta=2\). The data generating process is similar to that of the generalized normal distributions. First, we generate \(d\)-dimensional independent Beta(2, 2) variables. Then, we perturb these variables using a mean vector drawn from a unit normal distribution; for half of the samples, one, 10, or all variables are further perturbed by a normal distribution with variance \(a\). Finally, we introduce correlation by multiplying the data by a mixing matrix.

\begin{figure}[ht]
    \centering
    \includegraphics[width=0.99\linewidth]{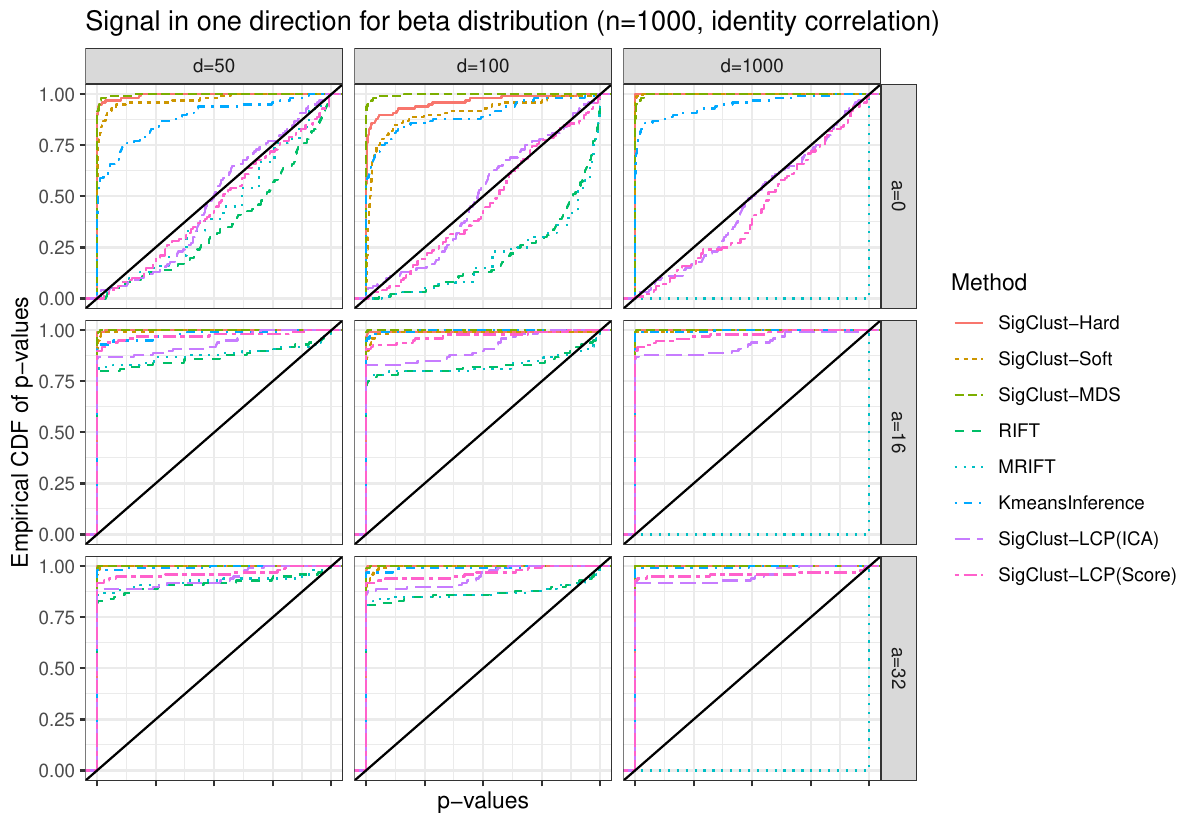}
    \caption{Simulation results for beta distributions. Here, $a$ denotes the distance between two clusters and $a=0$ denotes the null hypothesis of a single cluster.}
    \label{fig:sim_beta}
\end{figure}

Figure~\ref{fig:sim_beta} presents the results for \(n=1000\) with \(d\)-dimensional independent variables and a unidirectional signal. The performance of SigClust-LCP is similar to that observed under the generalized normal settings. Overall, SigClust-LCP consistently outperforms other methods in controlling the Type-I error and exhibits comparable power. Results for other settings can be found in the Supplementary Materials.

\subsection{Normal Distributions}

In the previous settings, the statistical power of SigClust-LCP relative to other methods was unclear because the other methods failed to control their Type-I error, naturally leading to inflated power. To address this issue, we conduct an analysis using normal distributions. The data-generating protocol follows that of the generalized normal distribution, with the only modification being that the tail parameter is set to \(\beta=2\).

\begin{figure}[ht]
    \centering
    \includegraphics[width=0.99\linewidth]{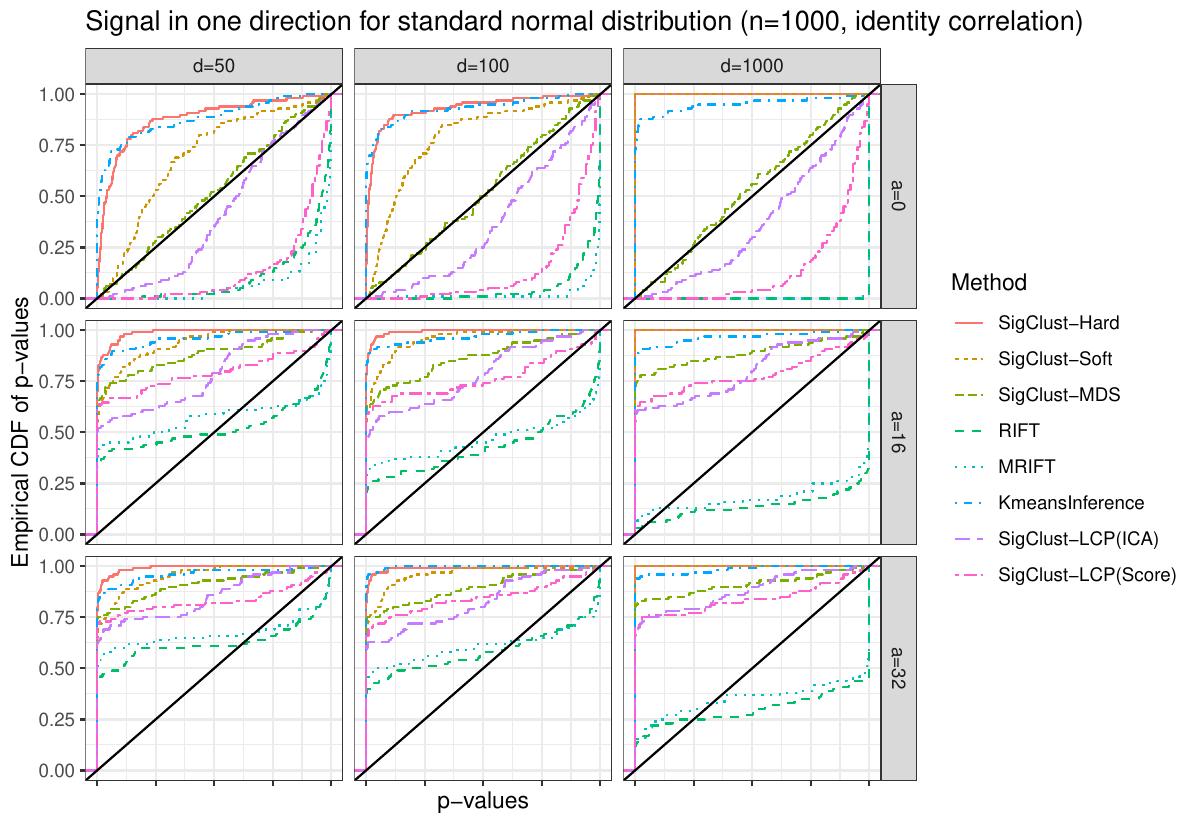}
    \caption{Simulation results for standard normal distributions. Here, $a$ denotes the distance between two clusters and $a=0$ denotes the null hypothesis of a single cluster.}
    \label{fig:sim_norm}
\end{figure}

Figure~\ref{fig:sim_norm} presents the results for standard normal distributions when \(n=1000\), with \(d\)-independent variables and a unidirectional signal. SigClust-LCP, along with SigClust-MDS methods, preserves the Type-I error. SigClust-Hard, SigClust-Soft, and $k$-means inference exhibit inflated Type-I error rates as the dimensionality increases due to inaccuracies in estimating the high-dimensional covariance matrix. In terms of statistical power, as expected, SigClust-LCP requires slightly stronger signal to reject the null hypothesis than SigClust-MDS. Notably, SigClust-LCP(Score) shows comparable power for moderate-dimensional settings, e.g., when $d=50$ and $100$. Since SigClust-LCP is nonparametric and considers a broader family of distributions than the original SigClust, this loss in power is not surprising and naturally creates a tradeoff in practice. However, it is important to note that verifying high-dimensional Gaussianity is usually challenging, which underscores the value of flexibility. Moreover, SigClust-LCP still outperforms RIFT and M-RIFT across all settings and exhibits acceptable power for practical use. Results for other scenarios are provided in the Supplementary Materials.

\subsection{Count Distributions}

Count distributions, such as the negative binomial and multinomial distributions, are commonly used to model gene expression data. We evaluate the performance of SigClust-LCP against recent methods for scRNA-seq data, including scSHC, CHOIR, and recall. With a sample size of \(n=1000\) and \(d=1000\), CHOIR exhibits anti-conservative behavior under the null for both distributions (Table \ref{tab:sc_rejection_rates_compact}). While scSHC and recall control the Type I error below 0.05 under the negative binomial distribution, they become anti-conservative under the multinomial distribution. In contrast, SigClust-LCP maintains Type I error control under both distributions. In terms of power, SigClust-LCP performs comparably to scSHC and recall, although it can be slightly conservative in certain scenarios. Additional results for varying dimensions and implementation details are provided in the Supplementary Materials.

\begin{table}[ht]
  \centering\scriptsize
  \setlength{\tabcolsep}{4pt}
  \renewcommand{\arraystretch}{0.95}
  \begin{tabular*}{\linewidth}{@{\extracolsep{\fill}} l *{8}{c}}
    \toprule
    & \multicolumn{4}{c}{Negative binomial} & \multicolumn{4}{c}{Multinomial} \\
    \cmidrule(lr){2-5}\cmidrule(lr){6-9}
    Method & $a=0$ & $a=0.25$ & $a=0.50$ & $a=0.75$ & $a=0$ & $a=0.25$ & $a=0.50$ & $a=0.75$ \\
    \midrule
    scSHC & 0.00 & 0.00 & 1.00 & 1.00 & 1.00 & 1.00 & 1.00 & 1.00 \\
    CHOIR & 0.90 & 0.95 & 1.00 & 1.00 & 0.14 & 0.18 & 0.99 & 1.00 \\
    Recall & 0.00 & 0.00 & 1.00 & 1.00 & 0.97 & 1.00 & 1.00 & 1.00 \\
    SigClust-LCP & 0.02 & 0.10 & 0.98 & 1.00 & 0.00 & 0.00 & 0.88 & 1.00 \\
    \bottomrule
  \end{tabular*}
  \caption{Empirical rejection rates at level 0.05 for the negative-binomial and multinomial simulations with $n=1000, d=1000$, based on 100 independent replications. The $a=0$ columns report Type-I error and the $a>0$ columns report power. Recall rejection means at least one selected feature at target FDR 0.05.}
  \label{tab:sc_rejection_rates_compact}
\end{table}

\subsection{Sensitivity Analysis}


SigClust-LCP involves several hyper-parameters, including the choice of truncation dimension and NN-associated parameters for score matching. To investigate the performance of SigClust-LCP with different truncation dimensions, we conduct experiments with the final dimension set to 2, 5, 10, and 30. Results show that SigClust-LCP is robust to the choice of truncation dimensions. Specifically, SigClust-LCP preserves the nominal Type-I error with all choices of truncation dimensions across different distributions.
We also investigate the performance of SigClust-LCP(Score) with different NN architectures and find that standard NN architectures generally lead to similar performance. Hence, we conclude that SigClust-LCP is robust for different hyper-parameters and include recommended hyper-parameters in the Supplementary Materials with detailed results. In addition, in the Supplementary Materials, we show that SigClust-LCP exhibits desirable statistical properties across different sample sizes and correlation structures, consistently outperforming other methods.

We also investigate the computational complexity of SigClust-LCP by evaluating its run time across different sample sizes and dimensions. We first compare our proposed score matching LCP to existing LCP estimators, such as \texttt{LogConcDEAD} \citep{culeLogConcDEADPackageMaximum2009} and ICA-based LCP, in terms of computational cost. Specifically, in low-dimensional settings, score matching LCP can estimate a 5-dimensional density within one minute, while it takes more than 20 minutes for \texttt{LogConcDEAD} to complete. Score matching LCP also shows its computational advantage in high-dimensional settings over ICA-based LCP. Next, we evaluate the run time in testing the clustering significance and find SigClust-LCP can complete in 1-5 minutes for a large variety of sample sizes and data dimensions. Details can be found in the Supplementary Materials. 

\section{Application to Single-cell RNA Sequencing Data}\label{sec:real}

A key use of clustering significance testing lies in classifying cell types from single-cell RNA sequencing data. In this work, we analyze the gene expression profiles of cells from \textit{Hydra}, a cnidarian polyp known for its extraordinary capacity to regenerate an entire organism from a small tissue piece. To probe the cellular diversity and regulatory programs underlying this regenerative ability, \cite{siebert2019stem} produced a dataset of 24,985 single-cell transcriptomes. \textit{Hydra} consists of three main cell lineages, endodermal epithelial, ectodermal epithelial, and interstitial, each sustained by its own pool of stem cells. The epithelial lineages, in particular, generate specialized structures along the body axis: the foot (including the basal disk and peduncle) at the aboral end, and the hypostome and tentacles at the oral end. Importantly, gene expression in these cells shifts dynamically with their spatial context. Here, we ask whether cells located at different positions along the body axis display distinct transcriptional signatures. To address this, we apply SigClust-LCP and several alternative methods to test clustering significance for (i) cells from the same annotated body axis within the same epithelial lineage, and (ii) cells drawn from different axes or different lineages. Notably, although single-cell RNA sequencing data are count-valued, discreteness does not violate the log-concave assumption -- for example, the Poisson distribution is log-concave. Since score-based LCP estimates a score function, the reference distribution in SigClust-LCP is taken as the continuous log-concave distribution that best approximates the observed data. In addition, we apply log-normalization and dimension reduction (see Supplementary Materials) to mitigate issues arising from discreteness.

Table \ref{tab:realnull} summarizes the clustering significance results for cells belonging to a single body-axis annotation from \cite{siebert2019stem}. SigClust-LCP, together with RIFT, MRIFT, and Recall, supports most of the original manual annotations, suggesting that cells within these annotated axes are largely homogeneous. In particular, these methods indicate that the body-column populations from the ectodermal and endodermal lineages may each contain more than one subpopulation, whereas the other annotated cell states are consistent with a single community. Notably, the \textit{Hydra} body column spans a broader range of body-axis positions than the other annotated body regions, and gene expression is known to vary with cell location \citep{siebert2019stem}. The significance results, with 7 out of 8 methods rejecting the single-cluster null for both body-column cell states, further support the possibility of additional positional subpopulations within the body column. Therefore, a more detailed subgroup analysis may reveal meaningful biological patterns. In contrast, several other methods fail to control the Type I error rate and tend to infer spurious subclusters. In particular, scSHC suggests that battery cells and tentacle cells may contain additional subpopulations, whereas CHOIR appears to overcluster the endodermal lineage. Other Gaussian-based methods, including SigClust-Hard, SigClust-Soft, and \(k\)-means inference, classify most annotated cell states as further clusterable. Visual inspection suggests that this inflated false-positive rate may arise from the non-Gaussian nature of the expression distributions.

To investigate the power of SigClust-LCP, we further apply the method to cell groups formed by combining two distinct lineages, whose biological separation is supported in the literature \citep{siebert2019stem}. Table \ref{tab:grouped_pairs} reports the clustering significance \(p\)-values of SigClust-LCP and competing methods for \textit{Hydra} cell pairs from different lineages, restricted to pairs for which at least one method yields a \(p\)-value greater than \(0.001\). We note that SigClust-Hard, SigClust-Soft, and \(k\)-means inference do not preserve Type I error under the single-cluster null; therefore, their \(p\)-values under the alternative are not directly valid for comparison and are included only for completeness. Among the 20 evaluated cell pairs, SigClust-LCP, MRIFT, and CHOIR reject all null hypotheses, providing strong evidence for the presence of multiple subpopulations. In contrast, RIFT and scSHC fail to reject two and four null hypotheses, respectively, suggesting that these methods may be conservative and could lead to under-clustering of genuine biological signals. Notably, scSHC and CHOIR also exhibit spurious clustering results under the single-cell-state null hypothesis. In addition, although Recall shows good Type I error preservation in the previous analysis, it rejects only 7 out of the 20 mixed-lineage cell groups, indicating that it may overlook genuine biological signals in practice. In summary, among the methods that preserve Type I error, the proposed SigClust-LCP is one of the most powerful: it rejects all alternative hypotheses in which cells are drawn from multiple lineages.

\begin{table}[ht]
  \centering\scriptsize
  \setlength{\tabcolsep}{2pt}
  \renewcommand{\arraystretch}{0.9}
  \begin{tabular*}{\linewidth}{@{\extracolsep{\fill}} l *{8}{c} }
    \toprule
    \makecell[l]{Cell Type}
      & \makecell[c]{SigClust-Hard}
      & \makecell[c]{SigClust-Soft}
      & \makecell[c]{$k$-means\\inference}
      & \makecell[c]{RIFT}
      & \makecell[c]{MRIFT}
      & \makecell[c]{scSHC}
      & \makecell[c]{CHOIR}
      & \makecell[c]{SigClust-\\LCP}\\
    \midrule
    {\bfseries Ectodermal}
      & \multicolumn{8}{c}{}\\
    Basal disk       & 0.034*    & 0.069      & 0.873   & 0.720   & 0.423    & 1.000       & 0.543       & 0.480\\
    Battery cell     & 0.068     & 0.041*     & $<$0.001* & 0.221 & 0.119    & $<$0.001*  & 0.157       & 0.180\\
    Head/Hypostome   & 0.131     & 0.124      & $<$0.001* & 0.999 & 0.942    & 1.000       & 0.812       & 1.000\\
    Peduncle         & 0.166     & 0.095      & 0.005*  & 0.755   & 0.332    & 0.970       & 0.127       & 0.960\\
    Body column      & 0.003*    & 0.005*     & $<$0.001* & 0.205 & 0.006*   & $<$0.001*  & $<$0.001*  & $<$0.001*\\
    \midrule
    {\bfseries Endodermal}
      & \multicolumn{8}{c}{}\\
    Foot             & 0.004*    & $<$0.001*  & $<$0.001* & 0.630 & 0.116    & 0.960       & 0.007*      & 0.580\\
    Head/Hypostome   & 0.040*    & 0.093      & $<$0.001* & 0.712 & 0.708    & 0.180       & 0.969       & 0.600\\
    Tentacle         & $<$0.001* & 0.007*     & 0.051   & 0.351   & 0.436    & $<$0.001*  & 0.024*      & 0.650\\
    Body column      & $<$0.001* & $<$0.001*  & 0.015*  & $<$0.001* & $<$0.001* & 1.000     & 0.004*      & 0.010*\\
    \bottomrule
  \end{tabular*}
  \caption{Real data application results on Hydra cells. Each row reports the $p$-value for a specific cell type, with an asterisk ($^*$) indicating statistical significance at the 0.05 level. Recall is omitted from the columns because it does not report pairwise cluster-level $p$-values; instead, it returns an FDR-controlled selected-gene count from the knockoff filter. For these same-cell-type comparisons, Recall selected no genes for all nine comparisons and did not reject at $q=0.05$.}
  \label{tab:realnull}
\end{table}

\begin{table}[ht]
  \centering\scriptsize
  \setlength{\tabcolsep}{3pt}
  \begin{tabular*}{\linewidth}{@{\extracolsep{\fill}} l *{5}{c}}
    \toprule
    \makecell[l]{Cell Types}
      & \makecell[c]{RIFT}
      & \makecell[c]{MRIFT}
      & \makecell[c]{scSHC}
      & \makecell[c]{CHOIR}
      & \makecell[c]{SigClust-\\LCP(Score)} \\
    \midrule
    \addlinespace[0.7ex]
    {\bfseries Ectodermal + Endodermal}
      & \multicolumn{5}{c}{}\\
    Head/Hypostome + Tentacle       & 0.166     & 0.017*    & $<$0.001* & $<$0.001* & $<$0.001* \\
    Peduncle + Tentacle             & 0.162     & 0.024*    & $<$0.001* & $<$0.001* & $<$0.001* \\
    Basal disk + Body column        & $<$0.001* & $<$0.001* & 0.520      & $<$0.001* & $<$0.001* \\
    Head/Hypostome + Body column    & $<$0.001* & $<$0.001* & 1.000      & $<$0.001* & $<$0.001* \\
    Peduncle + Body column          & $<$0.001* & $<$0.001* & 1.000      & $<$0.001* & $<$0.001* \\
    Body column + Head/Hypostome    & $<$0.001* & $<$0.001* & 0.360      & $<$0.001* & $<$0.001* \\
    \bottomrule
  \end{tabular*}
  \caption{Selected clustering significance $p$-values for Hydra cells from different body axis or cell lines. Rows are shown when at least one displayed method is not significant at the 0.05 level. Each entry reports the $p$-value for a mixture of cell types as assessed by the corresponding method, with an asterisk ($^*$) indicating statistical significance. Recall is omitted from the columns because it does not report pairwise cluster-level $p$-values; it rejects 7 out of 20 cross-lineage pairs.}
  \label{tab:grouped_pairs}
\end{table}

\section{Discussion}\label{sec:discussion}
{ In this work, we introduce a new nonparametric framework for evaluating the statistical significance of clustering. Motivated by single-cell RNA sequencing studies, where putative cell types or states are often defined by unsupervised clustering and then carried forward to downstream analysis, our approach leverages the LCP paradigm to construct the null distribution, thereby moving beyond the restrictive Gaussian assumptions that dominate existing methods. By accommodating a broad class of unimodal distributions, this framework substantially improves the relevance and robustness of clustering significance assessments in biomedical contexts, where data rarely follow Gaussian patterns.}

To support this framework, we design a score-matching algorithm specifically adapted to log-concave distributions. This enables accurate estimation of the null distribution in moderate to high dimensional spaces and broadens the applicability of current LCP solvers. Looking ahead, a natural extension of our work is to generalize the framework to multi-class clustering problems, including hierarchical and tree-structured settings. In addition, score-based LCP requires a score function and is therefore naturally suited to continuous distributions. Although our real-data analysis demonstrates robustness to gene expression counts, extending the framework to explicitly accommodate discrete data remains an important direction for future work.

\section*{Disclosure statement}
The authors declare no conflict of interest.

\section*{Data Availability Statement}
The genome data used for the \textit{Hydra} analysis are publicly available and can be downloaded from \url{https://research.nhgri.nih.gov/hydra/sequenceserver}.


\bibliographystyle{imsart-nameyear}
\bibliography{bibliography}
\end{document}